\documentclass[smus]{snow2e_m}
\usepackage{graphics}

\begin{document}

\def\D0{D\O}
\def\etmisv {\mbox{${\hbox{${\vec E}$\kern-0.5em\lower-.1ex\hbox{/}}}_T$}}
\def\etmis  {\mbox{${\hbox{$E$\kern-0.5em\lower-.1ex\hbox{/}}}_T$ }}
\def\pbarp  {\mbox{$\overline{p}p$} }
\def\pdf    {parton distribution function }
\def\pdfs   {parton distribution functions }
\def\ifmath#1{\relax\ifmmode #1\else $#1$\fi}%
\newcommand{\Afb}       {A_{\mathrm{FB}}}
\newcommand{\AFB}       {A_{\mathrm{FB}}}
\newcommand{\qqbar}     {\ifmath{\mathrm{q\bar{q}}}}
\newcommand{\uubar}     {\ifmath{\mathrm{u\bar{u}}}}
\newcommand{\ddbar}     {\ifmath{\mathrm{d\bar{d}}}}
\newcommand{\ppbar}     {\ifmath{\mathrm{p\bar{p}}}}

\begin{titlepage}

\headsep 2cm
\baselineskip 0.5cm

\evensidemargin 0.7cm

\begin{flushright}
{\large Fermilab-Conf-96/423 \\
UB-HET-96-06}
\end{flushright}

\vspace{1.5cm}

\begin{center}
{\Large \bf Precision Electroweak Physics at Future Collider 
Experiments}\footnote{\normalsize To appear in the {\it Proceedings of
the 1996 DPF/DPB Summer Study on New Directions for High-Energy
Physics}\\ 
(Snowmass~96), Snowmass, Colorado, June 25 -- July 12, 1996\\
Work supported in part by the U.S. Dept. of Energy
under contract DE-AC02-76CHO3000 and NSF grant PHY9600770}

\vspace{2cm}

{\Large U.~Baur\footnote{\normalsize Co-convener}} \\
\vspace{0.2cm}
{\large \it Physics Department, SUNY Buffalo, Buffalo, NY 14260, USA} \\
\vspace{1cm} 
{\Large M.~Demarteau$^2$} \\
\vspace{0.2cm}
{\large \it Fermi National Accelerator Laboratory, P.O.~Box 500,
Batavia, IL 60510, USA\\[15.mm]}

{\Large \bf Working Group Members}\\[3.mm]
{\large C.~Balazs {\large\it (MSU)}, D.~Errede {\large\it
(Urbana)}, S.~Errede {\it (Urbana)}, T.~Han {\it (Davis)}, \\[1.mm]      
S.~Keller {\large\it (FNAL)}, Y-K.~Kim {\large\it (Berkeley)}, 
A.V.~Kotwal {\large\it (Columbia)}, F.~Merritt {\large\it (Chicago)},
\\[1.mm] S.~Rajagopalan {\large\it (Stony Brook)}, 
R.~Sobey {\large\it (Davis)}, M.~Swartz {\large\it (SLAC)}, \\[1.mm]   
D.~Wackeroth {\large\it
(FNAL)}, J.~Womersley {\large\it (FNAL)}}

\vspace{3cm} 

{\Large \bf Abstract}\\[8.mm]
\begin{minipage}{6in}
\baselineskip 15pt
{\large We present an overview of the present status and prospects for
progress in electroweak measurements at future collider experiments 
leading to precision tests of the Standard Model of Electroweak
Interactions. Special attention is paid to the measurement of the $W$
mass, the effective weak mixing angle, and the determination of the top quark
mass. Their constraints on the Higgs boson mass are discussed.}
\end{minipage}
\end{center}

\end{titlepage}

\title{Precision Electroweak Physics at Future Collider 
Experiments\thanks{Work supported in part by the U.S. Dept. of Energy
under contract DE-AC02-76CHO3000 and NSF grant PHY9600770}}

\author{U.~Baur\thanks{Co-convener}\\ {\it Physics Department, SUNY 
Buffalo, Buffalo, NY 14260}\\[2.mm]
M.~Demarteau$^\S$\\ {\it Fermi National Accelerator Laboratory, P.O.
Box 500, Batavia, IL 60510}\\[4.mm]
{\bf Working Group Members}\\[3.mm]
C.~Balazs {\it (MSU)}, D.~Errede {\it
(Urbana)}, S.~Errede {\it (Urbana)}, T.~Han {\it (Davis)}, 
S.~Keller {\it (FNAL)}, \\[1.mm] Y-K.~Kim {\it (Berkeley)}, 
A.V.~Kotwal {\it (Columbia)}, F.~Merritt {\it (Chicago)}, 
S.~Rajagopalan {\it (Stony Brook)}, 
\\[1.mm] R.~Sobey {\it (Davis)}, M.~Swartz {\it (SLAC)}, D.~Wackeroth {\it
(FNAL)}, J.~Womersley {\it (FNAL)}}

\maketitle

%% Get rid of page numbering
%\thispagestyle{empty}\pagestyle{empty}
\thispagestyle{plain}\pagestyle{plain}

\begin{abstract} 
We present an overview of the present status and prospects for
progress in electroweak measurements at future collider experiments 
leading to precision tests of the Standard Model of Electroweak
Interactions. Special attention is paid to the measurement of the $W$
mass, the effective weak mixing angle, and the determination of the top quark
mass. Their constraints on the Higgs boson mass are discussed.
\end{abstract}

\section{Introduction}

The Standard Model (SM) of strong and electroweak interactions, based on the
gauge group $SU(3)_C\times SU(2)_L\times U(1)_Y$, has been extremely
successful phenomenologically. It has provided the theoretical
framework for the
description of a very rich phenomenology spanning a wide range of
energies, from the atomic scale up to the $Z$ boson mass, $M_Z$. It is being
tested at the level of a few tenths of a percent, both at very low
energies and at high energies~\cite{Marcel}, and has correctly predicted
the range of the top quark mass from loop corrections. However, the SM has a 
number of shortcomings. In particular, it does not explain the origin
of mass, the observed
hierarchical pattern of fermion masses, and why there are three
generations of quarks and leptons. It is widely believed that
at high energies, or in very high precision measurements, deviations
from the SM will appear, signaling the presence of new physics. 

In this report we discuss the prospects for precision tests of the
Standard Model at future collider experiments, focussing on electroweak
measurements. The goal of these measurements is to confront the SM
predictions with experiment, and to derive indirect information on the
mass of the Higgs boson. The existence of at least one Higgs boson is a
direct consequence of spontaneous symmetry breaking, the mechanism which
is responsible for generating mass of the $W$ and $Z$ bosons, and
fermions in the SM. 
In Section~II we identify some of the relevant parameters for precision
electroweak measurements, and
review the present experimental situation. Expectations from future
collider experiments are discussed in Section~III. We conclude with a
summary of our results. 

\section{Constraints on the Standard Model from Present Electroweak
Measurements}

There are three fundamental parameters measured with high precision
which play an important role as
input variables in Electroweak Physics. The fine structure constant,
$\alpha=1/137.0359895$ is known with a precision of $\Delta\alpha=0.
045$~ppm. The muon decay constant, $G_\mu=1.16639\times
10^{-5}$~GeV$^{-2}$ is measured with $\Delta G_\mu=17$~ppm from muon
decay~\cite{PDG}. Finally, the $Z$ boson mass, 
$M_Z=91.1863$~GeV/c$^2$~\cite{Marcel} is measured with
$\Delta M_Z=22$~ppm in experiments at LEP and SLC. Knowing these three
parameters, one can evaluate the $W$ mass, $M_W$, and the weak mixing
angle, $\sin^2\theta_W$, at tree level. When loop corrections are taken
into account, $M_W$ and $\sin^2\theta_W$ also depend on the top quark
mass, $M_t$, and the Higgs boson mass, $M_H$. The two parameters depend 
quadratically on $M_t$, and logarithmically on $M_H$. 

If the $W$ mass and the top quark mass are precisely measured, 
information on the mass of the Higgs boson can be extracted. 
Constraints on the Higgs boson mass can also be obtained from 
the effective weak mixing angle and $M_t$. The
ultimate test of the SM may lie in the comparison of these indirect 
determinations of $M_H$ with its direct observation at future colliders.

The mass of the top quark is presently determined by the CDF and D\O\ 
collaborations from $\bar tt$ production at the Tevatron in the
di-lepton, the lepton plus jets, and the all hadronic
channels~\cite{Gerdes}. The combined value of the top quark mass from
the lepton + jets channel, which yields the most precise result, is
\begin{equation}
M_t=175\pm 6~{\rm GeV/c}^2.
\end{equation}

The $W$ boson mass has been measured precisely by UA2, CDF, and D\O. 
Currently, the most accurate determination of $M_W$ comes from the
Tevatron CDF and D\O\ Run~Ia analyses~\cite{CDFWmass} and a preliminary D\O\ 
measurement~\cite{D0Wmass} based on data taken during Run~Ib. The
current world average is~\cite{Marcel}
\begin{equation}
M_W=80.356\pm 0.125~{\rm GeV/c}^2.
\label{mwdirect}
\end{equation}
Figure~\ref{FIG:ONE} compares the results of the current $M_W$ and $M_t$
measurements in the $(M_t,M_W)$ plane with those from indirect
measurements at LEP and SLC~\cite{Marcel}, and the SM prediction for 
different Higgs boson masses.
\begin{figure}[t]
\leavevmode
\begin{center}
\resizebox{8.9cm}{!}{%
\includegraphics{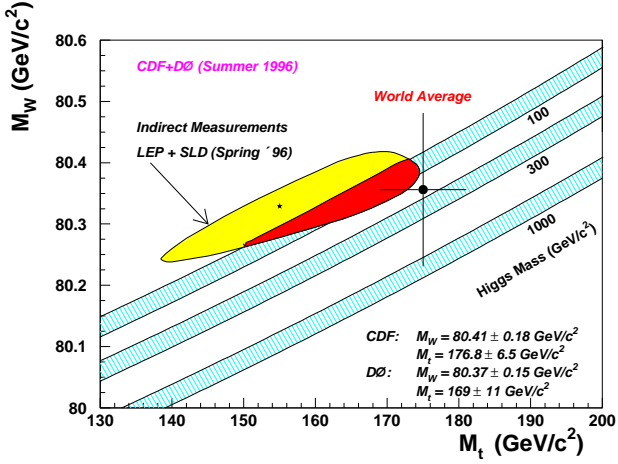}}
\end{center}
\caption{Comparison of the top quark and $W$ boson masses from current
direct and indirect measurements with the SM prediction.}
\label{FIG:ONE}
\end{figure}
The cross hatched bands show the SM prediction for the indicated Higgs
boson masses. The width of the bands is due primarily to the
uncertainty on the electromagnetic coupling constant at the $Z$ mass
scale, $\alpha(M_Z^2)$, which has been taken to be 
$\alpha^{-1}(M_Z^2)=128.89\pm 0.10$. Recent estimates 
give $\delta\alpha(M_Z^2)\approx 0.0004 - 0.0007$~\cite{Burk}, which
corresponds to $\delta\alpha^{-1}(M_Z^2)\approx 0.05 - 0.09$.

The uncertainty on $\alpha(M_Z^2)$ is dominated by the error on the
hadronic contribution to the QED vacuum polarization which originates
from the experimental error on the cross section for 
$e^+e^-\to {\rm hadrons}$. Using dispersion
relations~\cite{Cab}, the hadronic contribution to $\alpha(M_Z^2)$
can be related to the cross section of the process
$e^+e^-\to {\rm hadrons}$ via
\begin{equation}
\Delta\alpha_{\rm had}(M_Z^2)={\alpha M_Z^2\over 3\pi}~{\cal
P}\!\!\!\int_{4m_\pi^2}^\infty {R_{\rm had}(s')\over s'(s'-M_Z^2)}\, ds'\, ,
\end{equation}
where ${\cal P}$ denotes the principal value of the integral, and
\begin{equation}
R_{\rm had}={\sigma(e^+e^-\to {\rm hadrons})\over\sigma(e^+e^-\to
\mu^+\mu^-)}.
\end{equation}
The relative contributions to $\Delta\alpha_{\rm had}(M_Z^2)$ and the 
uncertainty are detailed in Fig.~\ref{FIG:ALPHA}~\cite{Burk}.
\begin{figure}[t]
\leavevmode
\begin{center}
\resizebox{8.9cm}{!}{%
\includegraphics{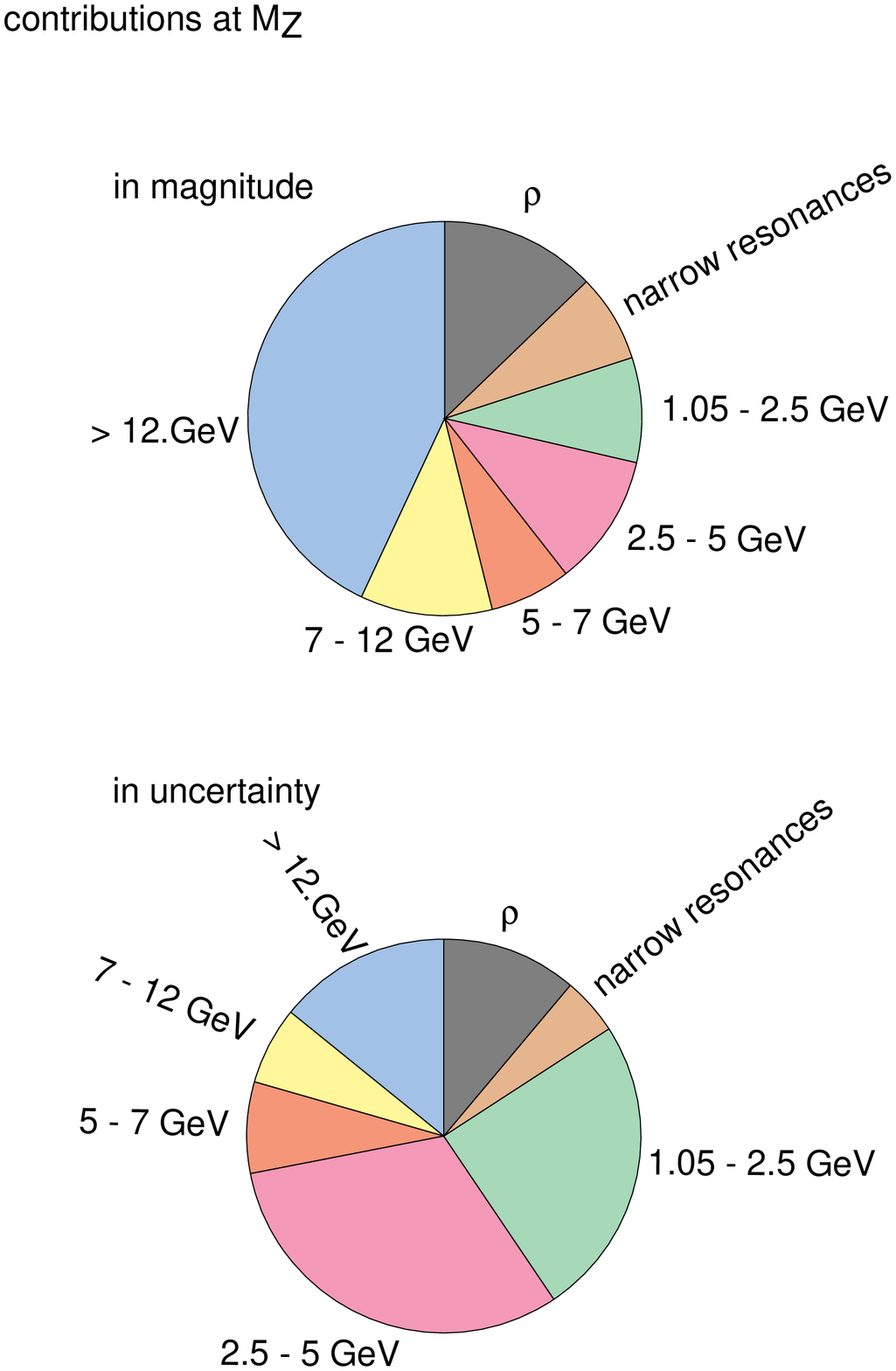}}
\end{center}
\caption{Relative contributions to $\Delta\alpha_{\rm had}(M_Z^2)$ in 
magnitude and uncertainty.}
\label{FIG:ALPHA}
\end{figure}
About 60\% of the uncertainty comes from the energy region between 1.05~GeV
and 5~GeV. More precise measurements of the total hadronic cross section
in this energy region, for example at Novosibirsk, DAP$\Phi$NE or BES 
may reduce the uncertainty on $\alpha(M_Z^2)$ by about a 
factor~2 in the near future. 

The $W$ mass can also be determined indirectly from radiative
corrections to electroweak observables at LEP and SLD, and from $\nu N$ 
scattering experiments. The current indirect value of $M_W$ obtained 
from $e^+e^-$ experiments, $M_W=80.337\pm 0.041^{+0.010}_{-0.
021}$~GeV/c$^2$~\cite{Marcel},
is in excellent agreement with the result obtained from direct
measurements (see Fig.~\ref{FIG:ONE}). The determination of $M_W$ from
$\nu N$ scattering will be discussed in Section~III.C.

The effective weak mixing angle, $\sin^2\theta^{lept}_{eff}$, has been 
determined with high precision from measurements of the forward backward
asymmetries at LEP, 
and the left-right asymmetries at the SLC~\cite{Marcel}. Here,
$\sin^2\theta^{lept}_{eff}$ is defined by 
\begin{equation}
\sin^2\theta^{lept}_{eff}={1\over 4}\left (1-{g_{V\ell}\over
g_{A\ell}}\right )\, ,
\end{equation}
where $g_{V\ell}$ and $g_{A\ell}$ are the effective vector and axial vector
coupling constants of the leptons to the $Z$ boson, and is related to
the weak mixing angle in the $\overline{\rm MS}$ scheme,
$\sin^2\hat\theta_W(M_Z)$, by~\cite{Sirlin}
\begin{equation}
\sin^2\theta^{lept}_{eff}\approx\sin^2\hat\theta_W(M_Z)+0.00028.
\end{equation}
A fit to the combined LEP and SLD asymmetry data yields
\begin{equation}
\sin^2\theta^{lept}_{eff}=0.23165\pm 0.00024.
\label{EQ:ASYM}
\end{equation}
The experimental constraints in the $(\sin^2\theta^{lept}_{eff},M_W)$
plane are compared with the SM predictions in Fig.~\ref{FIG:FOUR}.
\begin{figure}[t]
\leavevmode
\begin{center}
\resizebox{8.9cm}{!}{%
\includegraphics{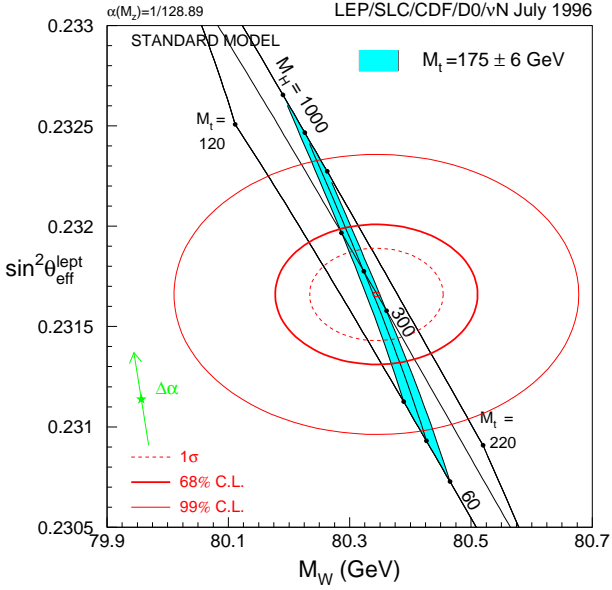}}
\end{center}
\caption{Comparison of $\sin^2\theta^{lept}_{eff}$ and the $W$ boson
mass from current direct and indirect measurements with the SM
prediction. The top quark and Higgs boson masses indicated in the figure
are all in GeV/c$^2$. }
\label{FIG:FOUR}
\end{figure}
The measured value of $\sin^2\theta^{lept}_{eff}$ agrees well with the
SM expectation. The star in the lower lefthand corner of 
Fig.~\ref{FIG:FOUR} indicates the $W$ mass and effective weak mixing
angle predicted by taking the running of $\alpha$ into account only. 
The arrow represents the current uncertainty on $M_W$ and the
effective weak mixing angle from $\Delta\alpha_{\rm had}(M_Z^2)$:
\begin{eqnarray}
\delta\sin^2\theta^{lept}_{eff}\bigl\vert_{\Delta\alpha}=0.00023, \\
\delta M_W\vert_{\Delta\alpha}= 12~{\rm MeV/c}^2.
\end{eqnarray}
The estimated theoretical error from higher orders introduces an
additional uncertainty of~\cite{alta}
\begin{eqnarray}
\delta\sin^2\theta^{lept}_{eff}\bigl\vert_{\rm th}=0.00008, \\
\delta M_W\vert_{\rm th}= 9~{\rm MeV/c}^2.
\end{eqnarray}

While direct measurements of $M_t$ and $M_W$
presently do not impose any constraints on the Higgs boson mass,
indirect measurements from LEP and SLD seem to indicate a preference for
a relatively light Higgs boson. The 68\% confidence level contours in
the $M_t$ and $M_H$ plane for the fits to LEP data only, and to all
data sets~\cite{Marcel} (LEP, SLD, CDF and D\O), are shown in 
Fig.~\ref{FIG:TWO}. Taking the theoretical error due to missing higher order
corrections into account, one obtains
\begin{equation}
M_H=149_{-82}^{+148}~{\rm GeV/c}^2,
\end{equation}
or
\begin{equation}
M_H<550~{\rm GeV/c}^2 \quad {\rm at\quad 95\%~CL}.
\end{equation}
\begin{figure}[t]
\leavevmode
\begin{center}
\resizebox{8.9cm}{!}{%
\includegraphics{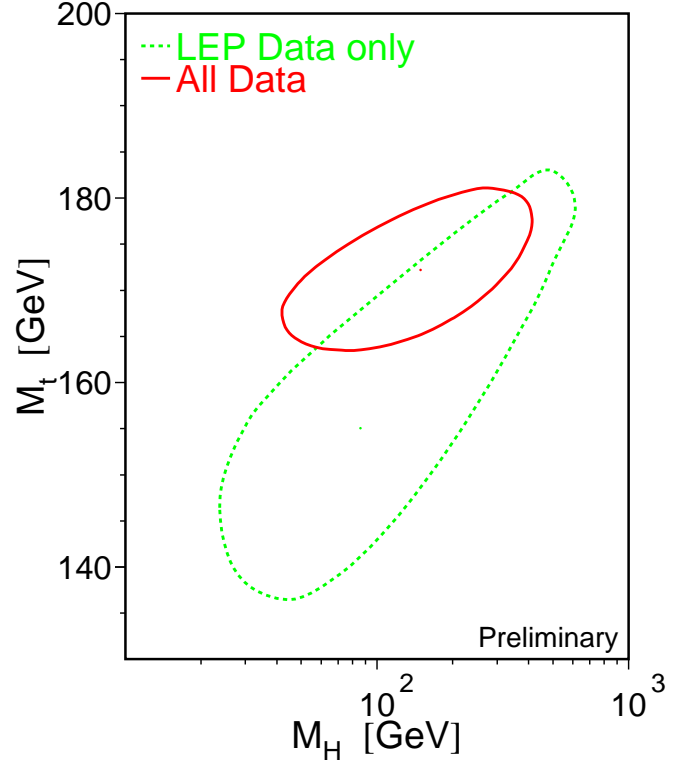}}
\end{center}
\caption{The 68\% confidence level contours in $M_t$ and $M_H$ for the
fits to LEP data only (dashed curve) and to all data (solid curve).}
\label{FIG:TWO}
\end{figure}
The results of such a fit from current data, however, should be 
interpreted with caution.
Removing one or two quantities from the fit can drastically change the
predicted Higgs boson mass range. Excluding from the fit the hadronic
width of the $Z$ boson, which depends on $\alpha_s$, results
in~\cite{Rosner} 
\begin{equation}
M_H=(560\times 1.5^{\pm 1})~{\rm GeV/c}^2.
\end{equation}
Omitting in addition the SLD data on
$A_{LR}$ which yield a somewhat low value for the effective weak mixing
angle, leads to $M_H=(820\times 1.7^{\pm 1})$~GeV/c$^2$. 

In the future, only marginal improvements of the indirect measurements
from LEP data are expected since LEP data taking at the $Z$ peak has ceased.
However, a significant reduction of the errors on $M_t$ and
$M_W$ from direct experiments at LEP2, the Tevatron (Run~I, Run~II and 
TeV33), the LHC, and perhaps the NLC and/or a $\mu^+\mu^-$ collider is
expected, which should result in a more stable prediction for $M_H$.
This will be discussed in more detail in the next Section.

\begin{figure}[th]
\leavevmode
\begin{center}
\resizebox{8.9cm}{!}{%
\includegraphics{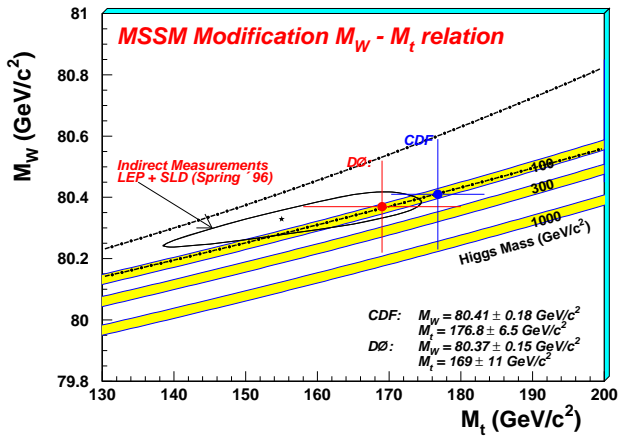}}
\end{center}
\caption{Predictions for $M_W$ as a function of $M_t$ in the SM (shaded
bands) and in the MSSM (area between the dot-dashed lines). The results
from direct CDF and D\O\ measurements, and from
indirect measurements at LEP and SLD are also shown.}
\label{FIG:THREE}
\end{figure}
Precise measurements of $M_W$ and $M_t$, if inconsistent with the range
allowed by the SM, could indicate the existence of new phenomena at or
above the electroweak scale, such as supersymmetry.
In the near future direct and indirect measurements of the top quark and
$W$ boson mass
are expected to begin to yield useful constraints on the parameter space of
the minimal supersymmetric standard model (MSSM). This is illustrated in
Fig.~\ref{FIG:THREE}, where the predictions for $M_W$ as a function of
$M_t$ in the SM (shaded bands) and in the MSSM (area between the dashed
lines) are shown, together with results from 
direct CDF and D\O\ measurements, and indirect measurements from LEP and 
SLD. The MSSM
band has been obtained by varying the model parameters so that they are
consistent with current experimental data. In addition, it
was assumed that no supersymmetric particles are found at
LEP2~\cite{SUSY}.

\section{High Precision Electroweak Physics at Future Colliders}

\subsection{Measurement of the Top Quark Mass}

The prospects of measuring the top quark mass in future collider
experiments are discussed in detail in Ref.~\cite{topgroup}. We
therefore only briefly summarize the results here. 

For the Tevatron, the expected accuracy in $M_t$ for Run~II ($\int\!{\cal
L}dt=2$~fb$^{-1}$) and for TeV33 ($\int\!{\cal L}dt=10-30$~fb$^{-1}$)
can be extrapolated using current and anticipated CDF and D\O\ 
acceptances and efficiencies, together with
theoretical predictions. Using various different methods and 
techniques~\cite{TEV2000},
one expects that $M_t$ can be determined to $\leq 4$~GeV/c$^2$ ($\leq 
2$~GeV/c$^2$) in Run~II (TeV33). The uncertainty on the top quark mass
will be dominated by systematic errors. Soft and hard gluon radiation,
and the jet transverse energy scale constitute the most important
sources of systematic errors in the top quark mass measurement at hadron
colliders. At the LHC, one also expects a
precision of about 2~GeV/c$^2$ for $M_t$~\cite{topgroup}. 

At an $e^+e^-$ Linear Collider (NLC) or a $\mu^+\mu^-$ collider,
the top quark
mass can be determined with very high precision from a threshold scan.
For an integrated luminosity of 10~fb$^{-1}$ (50~fb$^{-1}$), the
expected uncertainty on $M_t$ at the NLC is $\delta M_t\approx
500$~MeV/c$^2$ (200~MeV/c$^2$)~\cite{ZDR}. At a $\mu^+\mu^-$ collider,
the reduced beamstrahlung and initial state radiation result in a
better beam energy resolution which should make it possible to measure the
top quark mass with a somewhat higher precision than at the NLC, for
equal integrated luminosities. Simulations suggest $\delta
M_t\approx 300$~MeV/c$^2$ for 10~fb$^{-1}$~\cite{MC}. 

The precision which can be achieved for $M_t$ at different colliders is
summarized in Table~\ref{TAB:ONE}.
\begin{table}[th]
\begin{center}
\caption{Expected top quark mass precision at future colliders.}
\label{TAB:ONE}
\begin{tabular}{lc}
\hline
\hline
Collider & $\delta M_t$ \\
\hline
Tevatron (2~fb$^{-1}$) &  4~GeV/c$^2$\\
TeV33 (10~fb$^{-1}$) & 2~GeV/c$^2$ \\
LHC (10~fb$^{-1}$)  &   2~GeV/c$^2$ \\
NLC (10~fb$^{-1}$) & 0.5~GeV/c$^2$ \\
$\mu^+\mu^-$ (10~fb$^{-1}$) & 0.3~GeV/c$^2$ \\[1.mm]
\hline
\hline
\end{tabular}
\end{center}
\end{table}
In our subsequent calculations we shall always assume that the top quark
mass can be determined with a precision of
\begin{equation}
\delta M_t=2~{\rm GeV/c}^2.
\end{equation}

\subsection{Measurement of $\sin^2\theta^{lept}_{eff}$} 

\subsubsection{SLD}

Presently, the single most precise determination of the effective weak
mixing angle originates from the measurement of the left-right
asymmetry, 
\begin{equation}
A_{LR}={\sigma_L-\sigma_R\over \sigma_{tot}}
\end{equation}
at SLD. Here, $\sigma_{L(R)}$ is the total production cross section for
lefthanded (righthanded) electrons. In the SM, the left-right asymmetry
at the $Z$ pole, ignoring photon exchange contributions, is
related to the effective weak mixing angle by
\begin{equation}
A_{LR}={2\,(1-4\sin^2\theta^{lept}_{eff})\over
1+(1-4\sin^2\theta^{lept}_{eff})^2}~.
\end{equation}
If the planned luminosity upgrade~\cite{SLC2000} (``SLC2000'')
can be realized, it will be possible to collect $3\times 10^6$ $Z$
decays over a period of three to four years at SLD. This should result in an
uncertainty of
\begin{equation}
\delta\sin^2\theta^{lept}_{eff}=0.00012,
\end{equation}
which is approximately a factor~2 better than the current uncertainty from the
fit to the combined LEP and SLD asymmetry data (see 
Eq.~(\ref{EQ:ASYM})).

Further improvements could come from measurements of the left-right
forward-backward asymmetry in $e^+e^-\to\bar ff$, 
\begin{eqnarray}
\tilde A_{FB}^f(z)&=&{[\sigma_L^f(z)-\sigma_L^f(-z)]-[\sigma_R^f(z)
-\sigma_R^f(-z)] \over [\sigma_L^f(z)+\sigma_L^f(-z)]+[\sigma_R^f(z)
+\sigma_R^f(-z)]} \nonumber \\ 
&=& {2g_{Vf}g_{Af}\over g_{Vf}^2+g_{Af}^2}\,
{2z\over 1+z^2}\, ,
\end{eqnarray}
where $z=\cos\theta$, 
and $\theta$ is the scattering angle. $\tilde A_{FB}^f$ directly measures the 
coupling of the final state fermion $f$ to the $Z$ boson from which it is
straightforward to determine $\sin^2\theta^{lept}_{eff}$. In particular,
with the self-calibrating jet-charge technique~\cite{SLD}, a precise
measurement of the $Z\bar bb$ coupling should be possible.

\subsubsection{Hadron Colliders}

At hadron colliders, the forward backward asymmetry, $A_{FB}$, in
di-lepton production, $p\,p\hskip-7pt\hbox{$^{^{(\!-\!)
}}$}\to\ell^+\ell^-X$, ($\ell=e,\,\mu$),
makes it possible to measure the effective weak mixing angle. $A_{FB}$ is 
defined by 
\begin{equation}
A_{FB}={F-B\over F+B}\, ,
\label{EQ:DEFAFB}
\end{equation}
where
\begin{eqnarray}
F&=&\int_0^1{d\sigma\over d\cos\theta^*}\,d\cos\theta^*, \\[1.mm]
B&=&\int_{-1}^0{d\sigma\over d\cos\theta^*}\,d\cos\theta^*, 
\label{EQ:DEFFB}
\end{eqnarray}
and $\cos\theta^*$ is the 
angle between the lepton and the incoming quark in the $\ell^+\ell^-$
rest frame. In $p\bar p$ collisions at Tevatron energies, the flight
direction of the incoming quark to a good approximation coincides with
the proton beam direction. $\cos\theta^*$ can then be related to the
components of the lepton and anti-lepton four-momenta via~\cite{CS}
\begin{equation}
\cos\theta^*=2\,{p^+(\ell^-)p^-(\ell^+)-p^-(\ell^-)
p^+(\ell^+)\over m(\ell^+\ell^-)\sqrt{m^2(\ell^+\ell^-)
+p_T^2(\ell^+\ell^-)}} 
\label{EQ:CSTAR}
\end{equation}
with
\begin{equation}
p^\pm={1\over\sqrt{2}}\left (E\pm p_z\right ).
\end{equation}
Here, $m(\ell^+\ell^-)$ is the invariant mass of the lepton pair, $E$ is the
energy, and $p_z$ is the longitudinal component of the momentum vector.
In this definition of $\cos\theta^*$, the polar axis is defined to be 
the bisector of the
proton beam momentum and the negative of the anti-proton beam momentum
when they are boosted into the $\ell^+\ell^-$ rest frame. The four-momenta
of the quark and anti-quark cannot be determined individually. The 
definition of $\cos\theta^*$ in Eq.~(\ref{EQ:CSTAR})
has the advantage of minimizing the effects of the momentum ambiguity
induced by the parton transverse momentum.

First measurements of the effective weak mixing angle using the 
forward backward asymmetry at hadron colliders have been performed
by the UA1 and CDF collaborations~\cite{AFBCDF,AFBHI}. 
Figure~\ref{FIG:SIX}a shows the variation of $A_{FB}$ with the $e^+e^-$
invariant mass in $p\bar p\to e^+e^-$ for $\sqrt{s}=1.8$~TeV, assuming 
$\sin^2\theta^{lept}_{eff}=0.232$. The error
bars indicate the statistical errors for 100,000 events, corresponding
to an integrated luminosity of about 2~fb$^{-1}$. 
\begin{figure}[th]
\leavevmode
\begin{center}
\resizebox{8.9cm}{!}{%
\includegraphics{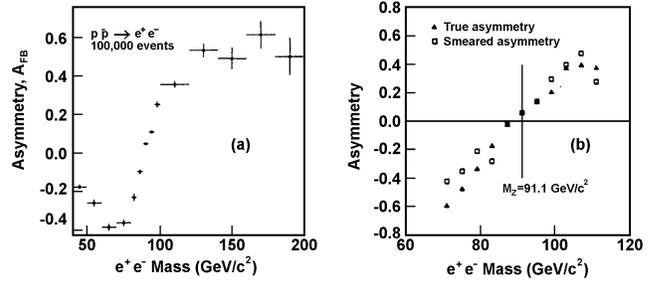}}
\end{center}
\caption{The forward backward asymmetry, $A_{FB}$, as a function of the
$e^+e^-$ invariant mass in $p\bar p\to e^+e^-$ events. (a) statistical
error for 100,000 events, corresponding to an integrated luminosity of
2~fb$^{-1}$ in an ideal detector; (b) including the effects of the D\O\
di-electron mass resolution.}
\label{FIG:SIX}
\end{figure}
The largest asymmetries occur at di-lepton invariant masses
of around 70~GeV/c$^2$ and above 110~GeV/c$^2$. A 
preliminary study of the systematic errors, indicates that most sources of
error are small compared with the statistical error. The main 
contribution to the systematic error originates from the uncertainty
in the parton distribution functions. Since the vector and axial vector 
couplings of $u$ and $d$ quarks to the $Z$ boson are different,
the measured asymmetry depends on the ratio of $u$ to  $d$
quarks in the proton. Most of the systematic errors are expected to
scale with $1/\sqrt{N}$, where $N$ is the number of events. 
The effect of the electromagnetic calorimeter resolution is rather
moderate, as shown in Fig.~\ref{FIG:SIX}b.
It is found that most of the sensitivity of this measurement to 
$\sin^2\theta_{eff}^{lept}$ is at $m(e^+e^-) \approx M_Z$ 
due to the strong variation of $A_{FB}$ with
$\sin^2\theta_{eff}^{lept}$ and the high statistics in this region.
Including QED radiative corrections, the $p\bar p\to e^+e^-$ forward 
backward asymmetry
in the $Z$ boson resonance region ($75~{\rm GeV/c}^2<m(e^+e^-)
<105~{\rm GeV/c}^2$) can be parameterized in terms of the effective weak 
mixing angle by~\cite{BKS}
\begin{equation}
A_{FB}=3.6\,(0.2464-\sin^2\theta^{lept}_{eff}). 
\end{equation}
The expected precision of $\sin^2\theta_{eff}^{lept}$ in the electron
channel (per experiment) versus the integrated luminosity at the
Tevatron is shown in
Fig.~\ref{FIG:SEVEN}, together with the combined current uncertainty
from LEP and SLD experiments. A similar precision is expected in the
muon channel.
\begin{figure}[th]
\leavevmode
\begin{center}
\resizebox{8.9cm}{!}{%
\includegraphics{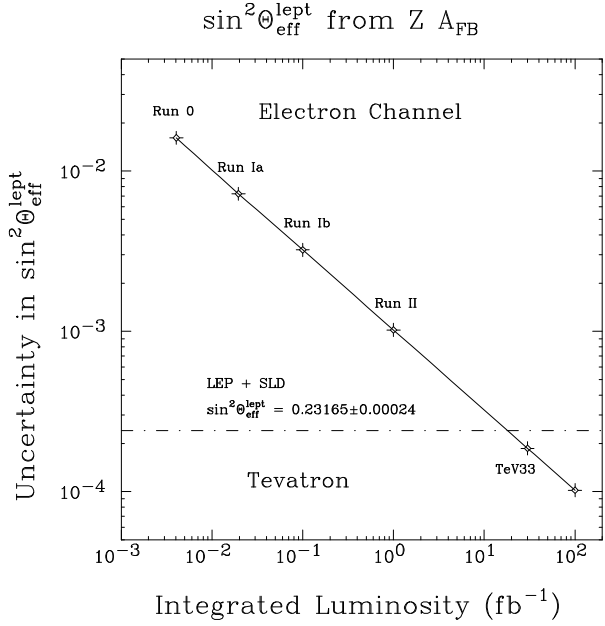}}
\end{center}
\caption{Projected uncertainty (per experiment) in 
$\sin^2\theta_{eff}^{lept}$ from the
measurement of $A_{FB}$ in the $Z$ pole region at the Tevatron versus
the integrated luminosity.}
\label{FIG:SEVEN}
\end{figure}
Combining the results of the electron and the muon channel, an overall 
uncertainty per experiment of 
\begin{equation}
\delta\sin^2\theta^{lept}_{eff}=0.00013
\end{equation}
is expected for an integrated luminosity of 30~fb$^{-1}$.

At the LHC, the lowest order 
$Z\to\ell^+\ell^-$ cross section is approximately 1.6~nb for each lepton
flavor. For the projected yearly integrated luminosity of
100~fb$^{-1}$, this results in a very large number of $Z\to\ell^+\ell^-$
events which, in principle, could be utilized to measure the forward
backward asymmetry and thus $\sin^2\theta^{lept}_{eff}$ with extremely
high precision~\cite{Fisher}. Since the original quark direction is 
unknown in $pp$ collisions, one
has to extract the angle between the lepton and the quark in the
$\ell^+\ell^-$ rest frame from the
boost direction of the di-lepton system with respect 
to the beam axis:
\begin{equation}
\cos\theta^*=2\,{|p_z(\ell^+\ell^-)|\over p_z(\ell^+\ell^-)}
{p^+(\ell^-)p^-(\ell^+)-p^-(\ell^-)p^+(\ell^+)\over 
m(\ell^+\ell^-)\sqrt{m^2(\ell^+\ell^-)
+p_T^2(\ell^+\ell^-)}}~.
\label{EQ:CSTAR1}
\end{equation}
in order to arrive at a non-zero forward-backward asymmetry.

In contrast to Tevatron energies, sea quark effects dominate at the LHC.
As a result, the probability, $f_q$, that the quark direction and the boost
direction of the di-lepton system coincide is significantly smaller 
than one. This considerably reduces the forward backward asymmetry.
Events with a large rapidity of the di-lepton system, $y(\ell^+\ell^-)$,
originate from collisions where at least one of the partons carries a
large fraction $x$ of the proton momentum. Since valence quarks
dominate at high values of $x$, a cut on the di-lepton rapidity 
increases $f_q$, and thus the asymmetry~\cite{Dittmar} and the 
sensitivity to the effective weak mixing angle. 

Imposing a $|y(\mu^+\mu^-)|>1$ cut and including QED corrections, 
the forward backward asymmetry at the LHC in the $\mu^+\mu^-$ channel 
in the $Z$ peak region ($75~{\rm GeV/c}^2<m(\mu^+\mu^-)<105~{\rm GeV/c}^2$)
can be parameterized by
\begin{equation}
A_{FB}=2.10\,(0.2466-\sin^2\theta^{lept}_{eff}) 
\end{equation}
for an ideal detector. For an integrated luminosity of 100~fb$^{-1}$, this 
then leads to an expected error of
\begin{equation}
\delta\sin^2\theta^{lept}_{eff}=4.5\times 10^{-5}.
\end{equation}
A similar precision should be achievable in the electron channel. 

However, electrons and muons can only be detected for pseudorapidities
$|\eta(\ell)|<2.4 - 3.0$ in the currently planned configurations of the
ATLAS~\cite{atlas} and CMS~\cite{cms} experiments at the LHC. The
finite pseudorapidity range available 
dramatically reduces the asymmetry. In the region around the
$Z$ pole, the asymmetry is again approximately a linear function of 
$\sin^2\theta^{lept}_{eff}$ with (for $\mu^+\mu^-$ final states)
\begin{equation}
A_{FB}=0.65\,(0.2488-\sin^2\theta^{lept}_{eff})~~ {\rm for}~~
|\eta(\mu)|<2.4.
\end{equation}
The finite rapidity coverage also
results in a reduction of the total $Z$ boson cross section by roughly a
factor~5. As a result, the uncertainty expected for 
$\sin^2\theta^{lept}_{eff}$ increases by almost a factor~7 to 
\begin{equation}
\delta\sin^2\theta^{lept}_{eff}=3.0\times 10^{-4}\qquad {\rm for} \qquad
|\eta(\mu)|<2.4.
\label{EQ:PREC1}
\end{equation}
In order to improve the precision beyond that expected from future SLC 
and Tevatron experiments, it will be necessary to detect electrons
and muons in the very forward pseudorapidity range, $|\eta|=3.0 - 5.0$,
at the LHC.

\subsubsection{NLC and $\mu^+\mu^-$ Collider}

The effective weak mixing angle can also be measured at the NLC in
fixed target M\o ller and Bhabha scattering. In fixed target
M\o ller scattering one hopes to
achieve a precision of $\delta\sin^2\theta^{lept}_{eff}=6\times
10^{-5}$~\cite{kumar}. In Bhabha scattering, it should be possible to
measure the effective weak mixing angle with a precision of 
a few~$\times 10^{-4}$~\cite{CG}, depending on the energy and polarization
available. Possibilities to determine the effective weak mixing angle at a
$\mu^+\mu^-$ collider have not been investigated so far.

\subsubsection{Constraints on $M_H$ from $\sin^2\theta^{lept}_{eff}$ and
$M_t$}

The potential of extracting useful information on the Higgs boson mass
from a fit to the SM radiative corrections and a precise measurement of 
$\sin^2\theta^{lept}_{eff}$ and $M_t$ is illustrated in 
Fig.~\ref{FIG:EIGHT}. Here we have assumed $M_t=176\pm 2$~GeV/c$^2$, 
$\sin^2\theta_{eff}^{lept}=0.23143\pm 0.00015$, and 
$\alpha^{-1}(M_Z^2)=128.89\pm 0.05$. 
\begin{figure}[th]
\leavevmode
\begin{center}
\resizebox{8.9cm}{!}{%
\includegraphics{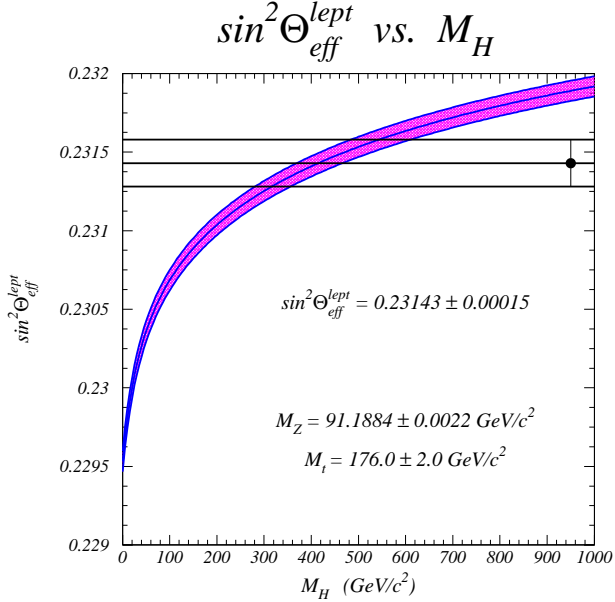}}
\end{center}
\caption{Predicted $\sin^2\theta_{eff}^{lept}$ versus the Higgs boson 
mass.}
\label{FIG:EIGHT}
\end{figure}
From such a measurement, one would find
$M_H=415^{+145}_{-105}$~GeV/c$^2$. The corresponding log-likelihood function
is shown in Fig.~\ref{FIG:NINE}.
\begin{figure}[th]
\leavevmode
\begin{center}
\resizebox{8.9cm}{!}{%
\includegraphics{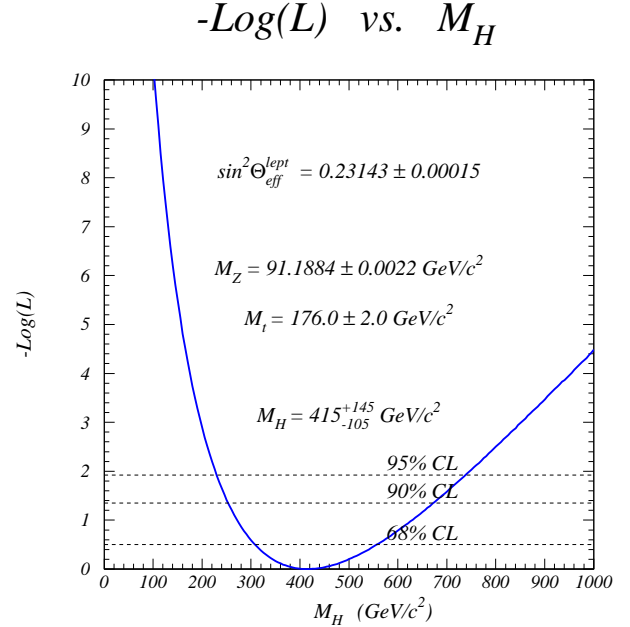}}
\end{center}
\caption{The negative log-likelihood function assuming
$\sin^2\theta_{eff}^{lept}=0.23143\pm 0.00015$ and $M_t=176\pm
2$~GeV/c$^2$.}
\label{FIG:NINE}
\end{figure}
From Fig.~\ref{FIG:EIGHT} it is obvious that the extracted Higgs boson
mass depends very sensitively on the central value of the effective weak
mixing angle. The relative error on the Higgs boson mass, $\delta
M_H/M_H\approx 30\%$, however, depends only on the uncertainty of higher
order corrections, $\sin^2\theta_{eff}^{lept}$, $M_t$, and $\alpha(M_Z^2)$. 
For the precision of $\sin^2\theta_{eff}^{lept}$ and $M_t$ assumed here,
the theoretical error from higher orders, and the uncertainty in
$\alpha(M_Z^2)$ begin to limit the accuracy which
can be achieved for the Higgs boson mass. 

\subsection{Precision Measurement of $M_W$ at Future Experiments}

\subsubsection{Deep Inelastic Scattering and HERA}

Future experiments provide a variety of opportunities to measure the mass of
the $W$ boson with high precision. In $\nu N$ scattering, $M_W$ can be
determined indirectly through a measurement of the neutral to charged 
current cross section ratio
\begin{equation}
R_\nu={\sigma(\nu N\to\nu X)\over\sigma(\nu N\to \mu^- X)}.
\end{equation}
In the SM, $R_\nu$ can be used to directly determine the weak mixing
angle via the lowest order expression
\begin{equation}
R_\nu={1\over 2}-\sin^2\theta_W+{5\over 9}\,(1+r)\sin^4\theta_W+C_\nu ,
\end{equation}
where
\begin{equation}
r={\sigma(\bar\nu N\to\mu^+ X)\over\sigma(\nu N\to\mu^- X)}\, ,
\end{equation}
and $C_\nu$ is a correction factor which incorporates, among others, 
effects due to charm production and longitudinal
structure functions. Electroweak radiative corrections modify
the leading order prediction. In the on-shell scheme, where
$\sin^2\theta_W=1-M^2_W/M_Z^2$ to all orders in perturbation theory,
the (leading) radiative corrections to $\sin^2\theta_W$ and $R_\nu$
almost perfectly cancel~\cite{stuart}. This implies that, in the SM,
$\nu N$ scattering directly measures the $W$ mass, given the very
precisely determined $Z$ boson mass. A new CCFR
measurement~\cite{farland} gives $M_W=80.46\pm 0.25$~GeV/c$^2$. With the
data which one hopes to collect in the NuTeV experiment during the
current Fermilab fixed target run, one expects~\cite{farland} 
\begin{equation}
\delta M_W\approx 100~{\rm MeV/c}^2.
\end{equation}
Figure~\ref{FIG:TEN} compares the
current results for $M_W$ from direct measurements at CDF, D\O\ and LEP2
(see below) with indirect determinations
from LEP and SLD via electroweak radiative corrections, and the $W$ mass
obtained from CCFR, other $\nu N$ experiments~\cite{nuN}, and the 
expectation for NuTeV. 
\begin{figure}[th]
\leavevmode
\begin{center}
\resizebox{8.9cm}{!}{%
\includegraphics{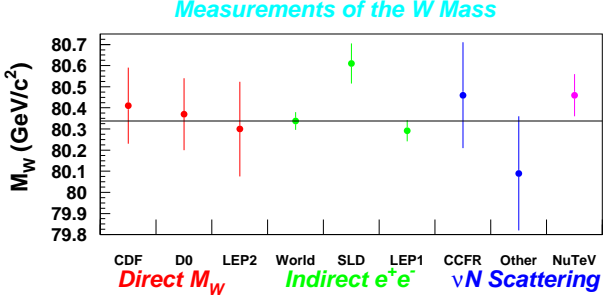}}
\end{center}
\caption{A comparison of direct and indirect measurements of the $W$
boson mass.}
\label{FIG:TEN}
\end{figure}

The $W$ mass can also be determined from 
measurements of the charged and neutral current cross sections at HERA.
Moving the low $\beta$ quadrupoles closer to the interaction region, one
hopes to achieve integrated luminosities of the order of
150~pb$^{-1}$ per year with a 70\% longitudinally polarized electron
beam. The expected constraints on $M_W$ and $M_t$, together with the
SM predictions for $M_H=100$~GeV/c$^2$ and $M_H=800$~GeV/c$^2$ are shown
in Fig.~\ref{FIG:ELEVEN}~\cite{hera}. When combined with a measurement of the
top quark mass with a precision of $\delta M_t=5$~GeV/c$^2$, the projected 
HERA results yield a precision of 
\begin{equation}
\delta M_W\approx 60~{\rm MeV/c}^2.
\label{EQ:HERA}
\end{equation}
Taking $\delta M_t=2$~GeV/c$^2$ instead only marginally improves the
accuracy on the $W$ mass. In deriving the result shown in 
Eq.~(\ref{EQ:HERA}), a 1\% relative systematic uncertainty of the charged 
and neutral current cross sections at HERA was assumed. For a systematic
error of 2\%, one finds $\delta M_W\approx 80~{\rm MeV/c}^2$.
\begin{figure}[t]
\leavevmode
\begin{center}
\resizebox{8.9cm}{!}{%
\includegraphics{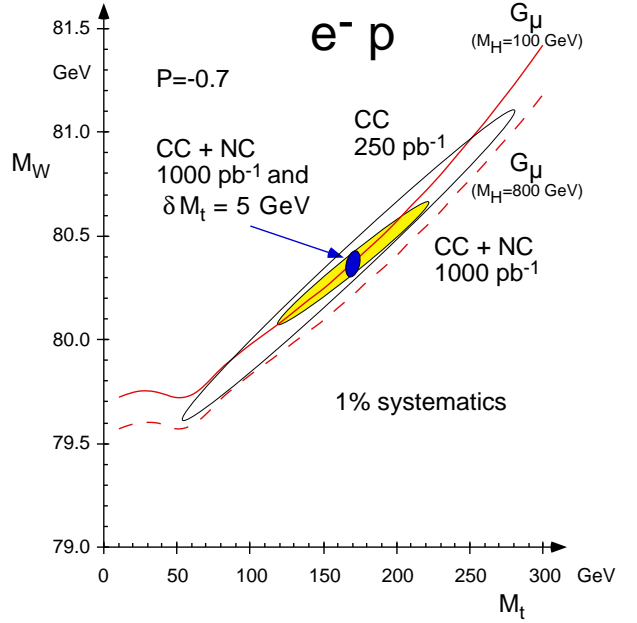}}
\end{center}
\caption{$1\sigma$ confidence contours in the $(M_W,M_t)$ plane from
polarized electron scattering at HERA (${\cal P}=-0.7$), utilizing 
charged current scattering alone for $\int\!{\cal L}dt=250$~pb$^{-1}$ (outer
ellipse), and neutral and charged current scattering for 
1~fb$^{-1}$ (shaded ellipse). Shown is also the combination of the
1~fb$^{-1}$ result with a direct top mass measurement with $\delta
M_t=5$~GeV/c$^2$ (full ellipse). The SM predictions are also shown for
two values of $M_H$ (from Ref.~[\ref{hera1}]).}
\label{FIG:ELEVEN}
\end{figure}

\subsubsection{LEP2 and NLC}

Precise measurements of the $W$ mass at LEP2~\cite{LEP2} can be obtained
using the enhanced statistical power of the rapidly varying total
$W^+W^-$ cross section at threshold~\cite{james}, and the sharp 
(Breit-Wigner) peaking behaviour of
the invariant mass distribution of the $W$ decay products. During the
recent LEP2 run at $\sqrt{s}=161$~GeV, the four LEP experiments have
each accumulated approximately 10~pb$^{-1}$ of data. The total $W^+W^-$ 
cross section as a function of the $W$ mass is shown in 
Fig.~\ref{FIG:TWELVE}, together with the preliminary experimental
result~\cite{watson}. Combining the
results obtained from the $W^+W^-\to jjjj$, the $W^+W^-\to\ell^\pm\nu
jj$ and the $W^+W^-\to\ell^+\nu\ell^-\nu$ ($\ell=e,\,\mu,\,\tau$)
channel, the $W$ pair production cross section at $\sqrt{s}=161$~GeV is
measured to be $\sigma(WW)=3.57\pm 0.46$~pb. This translates into a 
$W$ mass of~\cite{watson}
\begin{equation}
M_W=80.4\pm 0.2\pm 0.1~{\rm GeV/c}^2.
\end{equation}
\begin{figure}[t]
\leavevmode
\begin{center}
\resizebox{8.9cm}{!}{%
\includegraphics{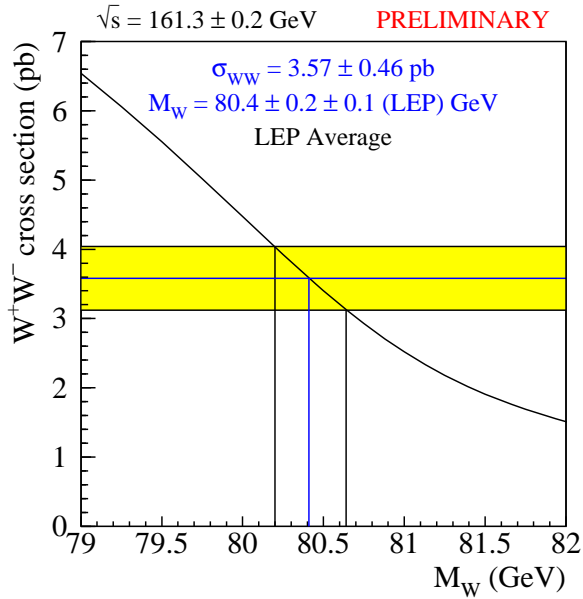}}
\end{center}
\caption{The total $W^+W^-$ cross section as a function of the $W$ boson
mass. The shaded band represents the cross section measured at LEP2.}
\label{FIG:TWELVE}
\end{figure}

A much more accurate measurement of $M_W$ will be possible in the future
through direct reconstruction methods when LEP2 will be running at 
energies well above the $W$ pair threshold. Here, the Breit-Wigner resonance
shape is directly reconstructed from the $W^\pm$ final states using
kinematic fitting techniques. The potentially most important limitation
in using this method originates from color reconnection~\cite{SK} and
Bose-Einstein correlations~\cite{SL} in the $W^+W^-\to jjjj$ channel. Taking
common errors into account, the expected overall precision from this
method at LEP2 for a total integrated luminosity of 500~pb$^{-1}$ per
experiment is anticipated to be~\cite{LEP2}
\begin{equation}
\delta M_W=35 - 45~{\rm MeV/c}^2.
\end{equation}

The same method can in principle also be used at the NLC. However, the 
beam energy spread limits the precision which one can hope to achieve at
an $e^+e^-$ Linear Collider.
Preliminary studies indicate that one can hope for a precision of 
$\delta M_W=20$~MeV/c$^2$ at best. No studies for a $\mu^+\mu^-$ collider have
been performed so far.

\subsubsection{Tevatron}

In $W$ events produced in a hadron collider in essence only two 
quantities are measured: the
lepton momentum and the transverse momentum of the recoil system. 
The latter consists of the ``hard'' $W$-recoil and the underlying event
contribution. For $W$-events these two are inseparable. 
The transverse momentum 
of the neutrino is then inferred from these two observables. Since the
longitudinal momentum of the neutrino cannot be determined unambiguously, 
the $W$-boson mass is usually extracted from the distribution in 
transverse:
\begin{equation} 
    M_T \,=\, \sqrt{2\, p_T(e) \, p_T(\nu) \, (1 - \cos\varphi^{e\nu}) },
\end{equation} 
where $\varphi^{e\nu}$ is the angle between the electron and neutrino in 
the transverse plane. The $M_T$ distribution sharply peaks at
$M_T\approx M_W$.

Both the transverse mass and lepton transverse momentum are, by
definition, invariant under longitudinal Lorentz boosts. In determining
the $W$ mass, 
the transverse mass is preferred over the lepton transverse 
momentum spectra because it is to first order independent of the 
transverse momentum of the $W$. Under transverse Lorentz boosts 
along a direction $\phi^*$, $M_T$ and $p_T(e)$ transform as
\begin{eqnarray*} 
    M_T^2 &\cong & {M_T^*}^2 \,-\, \beta^2 \, \cos^2\phi^* \, {M_L^*}^2
    \, ,    \\
    p_T(e) &\cong & p_T^*(e) \,+\, {1\over 2}\, p_T(W) \, \cos\phi^* \,
    ,
\end{eqnarray*} 
with $M_T^* = M_W \, \sin\theta^* $, $M_L^* = M_W \, \cos\theta^* $ and 
$\beta = p_T(W)/M_W$. The asterisk indicates quantities in the $W$ rest frame. 
The disadvantage of using the transverse mass is that it uses the 
neutrino transverse momentum which is a derived quantity. The neutrino 
transverse momentum is identified with the missing transverse energy in the 
event, which is given by 
\begin{eqnarray*}
    \etmisv \,=\, - {\displaystyle \sum_i } \, {\vec p}_{T_i}
            \,=\, -   {\vec p}_T(e) \,-\, {\vec p}_T^{\,rec} 
                \,-\, {\vec u}_T({\cal L}),
\end{eqnarray*}
where ${\vec p}_T^{\,rec}$ is the transverse momentum of the $W$-recoil 
system and ${\vec u}_T({\cal L})$ the transverse energy flow of the 
underlying event, which depends on the luminosity. 
It then follows that the magnitude of the missing 
$E_T$ vector and the true neutrino momentum are related as
\begin{equation} 
 \etmis \,=\, p_T(\nu) \,+\, {1\over 4} \, {u^2_T \over p_T(\nu)}\, .
\end{equation}
This relation can be interpreted as the definition of the neutrino
momentum scale. Note that the underlying event gives rise to a bias 
in the {\it measured } neutrino momentum with respect to the {\it true} 
neutrino momentum. In case there are more interactions per crossing, 
$| {\vec u}_T |$ behaves as a two-dimensional random walk and is 
proportional to $\sqrt{I_C}$, where $I_C$ is the average number of 
interactions per crossing. The shift in measured 
neutrino momentum is thus directly proportional to the number of 
interactions per crossing. The resolution increases as $\sqrt{I_C}$. 

The above equation for the missing transverse energy deserves some more
attention. The two components directly related to the $W$ decay, 
${\vec p}_T(e)$ and ${\vec p}_T^{\,rec}$, are only
indirectly affected by multiple interactions through the underlying event. 
It is the measurement of ${\vec u}_T({\cal L})$ which governs the 
luminosity dependence. 
Because of multiple interactions, ${\vec u}_T({\cal L})$ will 
show a dependence on luminosity following Poisson statistics, with the
two effects indicated above: $i)$ a degradation of the \etmis
resolution and $ii)$ a shift in the measured neutrino momentum. This is
demonstrated in Fig.~\ref{FIG:THIRTEEN} where we show the $M_T$
distribution for various values of $I_C$ at the Tevatron. For Run~II
one expects $I_C\approx 3$, and at TeV33, $I_C\approx
6-9$~\cite{GJ}.
\begin{figure}[t]
\leavevmode
\begin{center}
\resizebox{8.9cm}{!}{%
\includegraphics{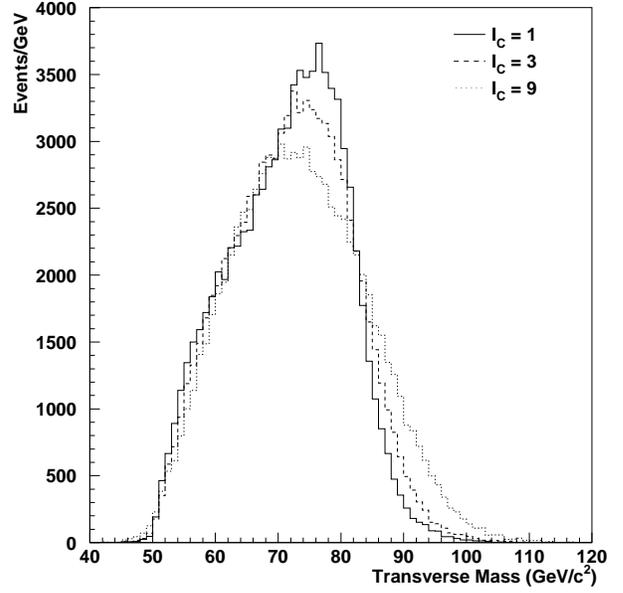}}
\end{center}
\caption{The effect of multiple interactions on the $W$ transverse mass 
distribution at the Tevatron. Standard kinematic cuts of 
$p_T(e)>25$~GeV/c, $|\eta(e)|<1.2$, 
$\etmis >25$~GeV and $p_T(W)<30$~GeV/c are imposed. The effect of
multiple interactions is simulated by adding additional minimum bias
events to the event containing the $W$ boson.}
\label{FIG:THIRTEEN}
\end{figure}
Both effects, of course, propagate into the measurement of the 
transverse mass and the uncertainty on $M_W$ will not follow the simple 
$1/\sqrt{N}$ rule anymore~\cite{tev20001}. 
In addition, however, the detector response to high luminosities needs
to be folded in. In the above discussion it was assumed that the detector
response is linear to the number of multiple interactions which  
in general is not the case. The effects of pile-up in the calorimeter 
and occupancy in the tracking detectors produce a $\sim 7\%$ shift
in $p_T$ for an electron with transverse momentum of 40~GeV/c at ${\cal
L}=10^{33}~{\rm cm}^{-2}~{\rm s}^{-1}$, which 
will further affect the uncertainty on the $W$ mass
adversely~\cite{ashtosh}.

Another uncertainty that will not, and has not in the past, scaled 
with luminosity is the theoretical uncertainty coming from the 
$p_T(W)$ model and the uncertainty on the proton structure. 
Parton distributions and the spectrum in $p_T(W)$ are correlated. 
The \D0 experiment has addressed this correlation in the determination
of its uncertainty on the $W$ mass~\cite{CDFWmass,D0Wmass}. The parton 
distribution functions are
constrained by varying the CDF measured $W$ charge asymmetry within the 
measurement errors, while at the same time utilizing all the available data. 
New parametrizations of the
CTEQ~3M parton distribution function were obtained that included in the
fit the CDF $W$ asymmetry data from Run~Ia~\cite{cdf_wasym}, 
where all data points had
been moved coherently up or down by one standard deviation. 
In addition one of the parameters, which describes the
$Q^2$-dependence of the parameterization of the non-perturbative
functions describing the $p_T(W)$ spectrum~\cite{ly}, was varied. 
The constraint on this parameter was provided by the measurement of 
the $p_T(Z)$ spectrum. 
The uncertainty due to parton distribution functions and the 
$p_T(W)$ input spectrum was then assessed by varying simultaneously 
the parton distribution function, as determined by varying
the measured $W$ charge asymmetry, and the parameter describing the 
non-perturbative part of the $p_T(W)$ spectrum. 

The CDF experiment uses their measurement of the $W$ charge asymmetry 
as the sole constraint on the uncertainty due to the parton
distribution functions.
Figure~\ref{FIG:FOURTEEN} compares the preliminary CDF $W$ charge
asymmetry measurement~\cite{Arie} with several recent fits to parton
distribution functions.
\begin{figure}[t]
\leavevmode
\begin{center}
\resizebox{8.9cm}{!}{%
\includegraphics{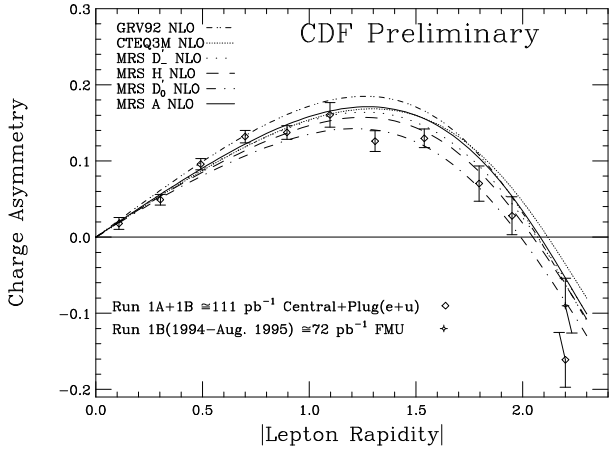}}
\end{center}
\caption{Comparison of the CDF $W$ asymmetry measurement with recent
NLO \pdf predictions. }
\label{FIG:FOURTEEN}
\end{figure}
Figure~\ref{FIG:FIFTEEN} shows the correlation between the uncertainty on
the $W$ mass, $\Delta M_W$, and
\begin{equation}
\Delta\sigma(A(\eta))={\langle A_{PDF}(\eta)\rangle-\langle
A_{data}(\eta)\rangle\over\delta A_{data}(\eta)}\, ,
\end{equation}
the deviation between the average
measured asymmetry for Run~Ia and~Ib data and various recent NLO \pdf
fits~\cite{Arie}. 
\begin{figure}[t]
\leavevmode
\begin{center}
\resizebox{8.9cm}{!}{%
\includegraphics{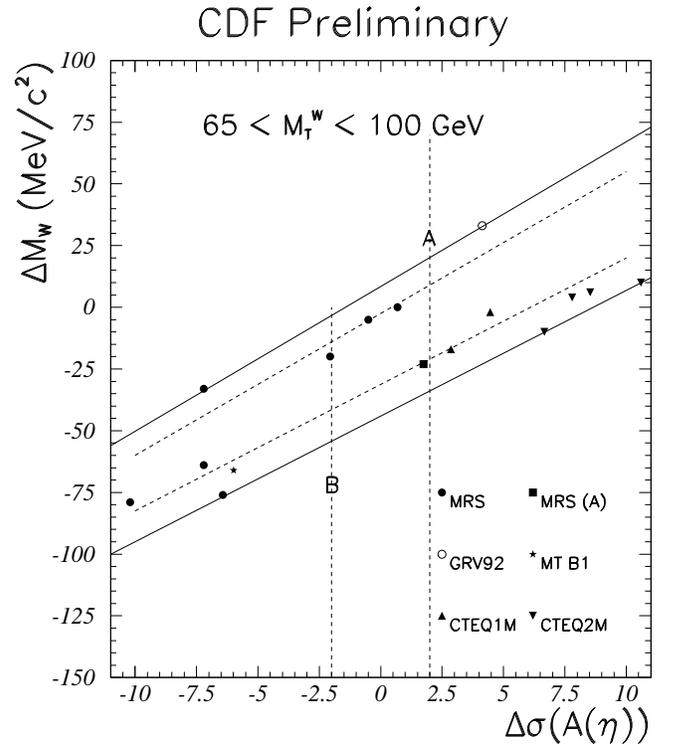}}
\end{center}
\caption{The correlation between the uncertainty in the $W$ mass and the
deviation between the average measured asymmetry for Run~Ia and~Ib CDF
data for several recent \pdfs.}
\label{FIG:FIFTEEN}
\end{figure}
The fitted $W$ mass is seen to be strongly correlated 
with the $W$ charge asymmetry. The $W$ charge asymmetry, however, is mainly 
sensitive to the slope of the ratio of the $u$ and $d$ quark \pdfs 
\begin{equation} 
    A(y_W) \,\propto\,
            { d(x_2) \,/\, u(x_2)  \,-\, d(x_1) \,/\, u(x_1)  \over 
              d(x_2) \,/\, u(x_2)  \,+\, d(x_1) \,/\, u(x_1)  }
\end{equation} 
and does not probe the full parameter range describing the \pdfs. 

Future measurements of the $p_T(Z)$ distribution will provide a
constraint on the $p_T$ distribution of the $W$ boson. Moreover, 
the measurements of the $W$ charge asymmetry, together with measurements 
from deep inelastic scattering experiments, will provide further
constraints on the parton distribution functions. An effort needs to be
made, though, to provide the experiments with parton distributions with
associated uncertainties. 

At high luminosities alternate methods to determine the $W$-mass
may be advantageous. Because of the similarity of $W$ 
and $Z$ production, methods based on ratios of relevant quantities, such
as the charged lepton transverse momenta are particularly 
interesting~\cite{KG,srini}. The ratio of the lepton $p_T$ distributions
is thought to be very promising for fitting the $W$ mass in
the high luminosity regime since the procedure is independent of many
resolution effects. However, the shapes of the lepton transverse
momentum distributions are sensitive to the differences in the $W$ and
$Z$ production mechanisms, which need to be better understood.

Here we concentrate on a similar method which
utilizes the transverse mass ratio of $W$ and $Z$ bosons~\cite{srini}.
Preliminary results from an
analysis of the transverse mass ratio have recently been presented by
the D\O\ Collaboration~\cite{trmr}. Only the electron channel will be
discussed in the following, although the method is expected to work for
muon final states as well.

The transverse mass ratio method treats the $Z\to e^+e^-$ sample similar
to the $W\to e\nu$ sample, thus cancelling many of the common systematic
uncertainties. A transverse mass for the $Z$ boson is constructed with 
one of the decay electrons, while the $\etmis$ is derived by adding
the transverse energy of the other electron to the residual $\etmis$ in the
event. Hence, two such combinations can be formed for each $Z$ event.

The $Z$ transverse mass distribution is scaled down in finite steps and
compared with the $M_T$ distribution of the $W$ boson. The $W$ mass is
then determined from the scale factor $(M_W/M_Z)$ which gives the best
agreement between the $M_T$ distributions using a Kolmogorov test.
Since differences
in the production mechanism, acceptances and resolution effects between
the $W$ and the $Z$ sample lead to differences in the shapes of the
transverse mass distributions, one has to correct for these effects.

The dominant systematic uncertainty arises from the uncertainty on the
underlying event. 
Electromagnetic and hadronic resolution effects mostly cancel in the
transverse mass ratio, as expected. The systematic uncertainty due to 
the \pdfs and
the transverse momentum of the $W$ boson is reduced by more than a
factor~3 compared with that found using the conventional $W$ transverse 
mass method~\cite{CDFWmass}. The total systematic error from the D\O\
Run~Ia data sample is estimated to be 75~MeV/c$^2$. For comparison, the
total systematic error obtained using the transverse mass distribution
of the $W$ using D\O\ Run~Ia data is 165~MeV/c$^2$~\cite{CDFWmass}. 

In the analysis of the Run~Ia data sample, electrons from $W$ and $Z$
decay are identified as in the conventional $W$ mass analysis. $W$
candidates are selected by requiring $p_T(e)>30$~GeV/c and $p_T(\nu)
>30$~GeV/c, while electrons from $Z$ decays are required to have $p_T(e)
>34$~GeV/c, since they are eventually scaled down. Electrons from $W$
decay and at least one electron from $Z$ decay are required to be in the
central pseudorapidity region, $|\eta(e)|<1.1$. $Z$ events are used
twice if both electrons fall in the central region. The shape comparison
is performed in the fitting window $65~{\rm GeV/c}^2<M_T<100~{\rm
GeV/c}^2$. The
selected $Z$ sample is scaled down in finite steps and, at every step,
the shape of the $Z$ and $W$ $M_T$ distribution is compared using the
Kolmogorov test. Figure~\ref{FIG:SIXTEEN} shows the $M_T(Z)$ distribution
superimposed on the $M_T(W)$ distribution for one of the fits.
\begin{figure}[t]
\leavevmode
\begin{center}
\resizebox{8.9cm}{!}{%
\includegraphics{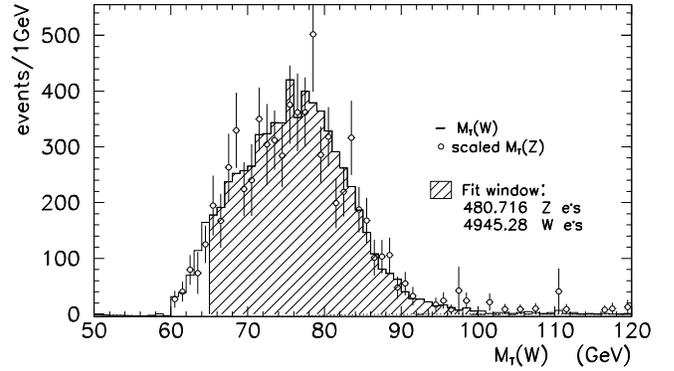}}
\end{center}
\caption{The Run~Ia D\O\ $M_T(W)$ distribution (histogram) with the
scaled $M_T(Z)$ distribution (points) superimposed. }
\label{FIG:SIXTEEN}
\end{figure}
The preliminary result for $M_W$ from Run~Ia data is 
\begin{equation}
M_W=80.160\pm 0.360({\rm stat})\pm 0.075({\rm syst})~{\rm GeV/c}^2.
\label{mtresult}
\end{equation}
The limitation of the method described here comes entirely from the
limited $Z$ statistics, which is expected to scale exactly as 
$1/\sqrt{N}$ in future experiments. 

The power of the $M_T$
ratio method becomes apparent when one compares the uncertainty on $M_W$
expected for 1~fb$^{-1}$ and 10~fb$^{-1}$ with that expected from the
traditional $W$ transverse mass analysis~\cite{tev20001}. The results
for both methods are listed in Table~\ref{TAB:TWO}. To calculate the
projected statistical (systematic) errors in the transverse mass ratio
method, we have taken the errors of Eq.~(\ref{mtresult}) and scaled them
with $1/\sqrt{N}$ ($\sqrt{I_C/N}$), assuming $I_C=3$ ($I_C=9$) for
1~fb$^{-1}$ (10~fb$^{-1}$). Both, electron and muon channels are
combined in Table~\ref{TAB:TWO}, assuming that the two channels yield 
the same precision in $M_W$. 
\begin{table}[th]
\begin{center}
\caption{Projected statistical and systematic errors (per experiment) on
the $W$ mass at the Tevatron, combining the $W\to e\nu$ and $W\to\mu\nu$
channel.}
\label{TAB:TWO}
\begin{tabular}{lcc}
\hline
\hline
\multicolumn{3}{c}{traditional $M_T$ analysis}\\
\hline
 & $\int\!{\cal L}dt=1~{\rm fb}^{-1}$ & 
$\int\!{\cal L}dt=10~{\rm fb}^{-1}$ \\
$\delta M_W$ & $I_C=3$ & $I_C=9$ \\
\hline
statistical & $29~{\rm MeV/c}^2$ & $17~{\rm MeV/c}^2$ \\
systematic & $42~{\rm MeV/c}^2$ & $23~{\rm MeV/c}^2$ \\
\hline
total & $51~{\rm MeV/c}^2$ & $29~{\rm MeV/c}^2$ \\
\hline
\hline
\multicolumn{3}{c}{$W/Z$ transverse mass ratio}\\
\hline
 & $\int\!{\cal L}dt=1~{\rm fb}^{-1}$ & 
$\int\!{\cal L}dt=10~{\rm fb}^{-1}$ \\
$\delta M_W$ & $I_C=3$ & $I_C=9$ \\
\hline
statistical & $29~{\rm MeV/c}^2$ & $9~{\rm MeV/c}^2$ \\
systematic & $10~{\rm MeV/c}^2$ & $6~{\rm MeV/c}^2$ \\
\hline
total & $31~{\rm MeV/c}^2$ & $11~{\rm MeV/c}^2$ \\
\hline
\hline
\end{tabular}
\end{center}
\end{table}

The $W$ mass can also be determined from the transverse energy
(momentum) distribution of the electron (muon) in $W\to e\nu_e$
($W\to\mu\nu_\mu$) events, which peaks at $M_W/2$. The prospects of a 
precise measurement of $M_W$ from the $E_T(e)$ distribution in Run~II 
and at TeV33 have been investigated in
Ref.~\cite{ashtosh}. The measurement of the lepton four-momentum
vector is independent of the $\etmis$ resolution, and the electron $E_T$
resolution is dominated by the intrinsic calorimeter resolution. Hence
the statistical uncertainty of the $W$ mass measurement from the $E_T(e)
$ distribution is expected to scale approximately as $1/\sqrt{N}$. 
Simulations have shown that a
sample of 30,000 events (similar to the D\O\ Run~Ib data sample) gives a
statistical error on the $W$ mass of 100~MeV/c$^2$ from the $E_T(e)$
fit. This is in agreement with the result of the preliminary D\O\
Run~Ib $W$ mass analysis~\cite{Rijs}. The systematic error from this
method is expected to be about 170~MeV/c$^2$ for the same
number of events. Scaling the total uncertainty as $1/\sqrt{N}$, the
projected uncertainty of $M_W$ from the electron $E_T$ fit is:
\begin{eqnarray}
\delta M_W = 55~{\rm MeV/c}^2 & {\rm for} & 1~{\rm fb}^{-1}, \nonumber \\
\delta M_W = 18~{\rm MeV/c}^2 & {\rm for} & 10~{\rm fb}^{-1}.
\label{EQ:ET}
\end{eqnarray}

In estimating the uncertainties given in Eq.~(\ref{EQ:ET}) and 
Table~\ref{TAB:TWO}, we have assumed that the current uncertainty from 
\pdfs and the theoretical uncertainty originating from higher
order electroweak corrections can be drastically reduced in 
the future. In order to measure $M_W$ with high precision, it is crucial 
to fully control higher order electroweak (EW) corrections. So far, only 
the final state ${\cal O}(\alpha)$ photonic corrections have been 
calculated~\cite{BK}, using an 
approximation which indirectly estimates the soft + virtual part
from the inclusive ${\cal O}(\alpha^2)$ $W\to\ell\nu(\gamma)$
width and the hard photon bremsstrahlung contribution. Using this
approximation, electroweak corrections were found to shift the $W$ mass
by about $-65$~MeV/c$^2$ in the electron, and $-170$~MeV/c$^2$ in the
muon channel~\cite{CDFWmass,D0Wmass}. 

Currently, a more complete calculation of the ${\cal O}(\alpha)$ EW 
corrections, which takes
into account initial and final state corrections, is being carried
out~\cite{BKW}. The calculation is performed using standard Monte Carlo 
phase space slicing techniques for NLO calculations. In calculating the 
initial state radiative corrections, mass (collinear) singularities are
absorbed into the parton distribution functions through 
factorization, in complete analogy to the QCD case. QED corrections to
the evolution of the \pdf are not taken into account. A study of the 
effect of QED on the evolution indicates that
the change in the scale dependence of the PDF is small~\cite{spies}.
To treat the QED radiative corrections in a consistent way, they
should be incorporated in the global fitting of the PDF. The
relative size and the characteristics of the various contributions
to the EW corrections to $W$ production is shown in Fig.~\ref{fig:ratio}. 
\begin{figure}[t]
\leavevmode
\begin{center}
\resizebox{8.9cm}{!}{%
\includegraphics{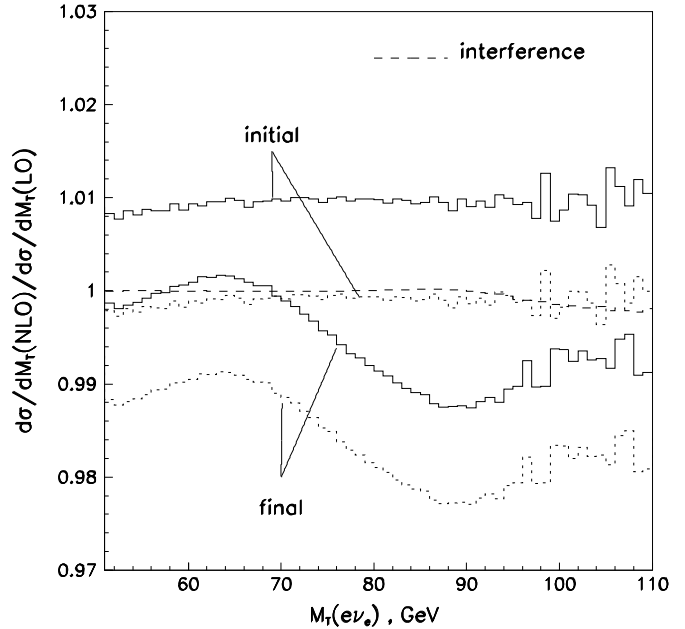}}
\end{center}
\caption{The ratio of the NLO to LO $M_T(e\nu_e)$ distribution for
various individual contributions:~the
QED-like initial or final state contributions (solid), the
complete ${\cal O}(\alpha)$ initial and final state 
contributions (short dashed) and the initial--final
state interference contribution (long dashed).}
\label{fig:ratio}
\end{figure}

Initial state (photon and weak) radiative corrections are found to be uniform
and, therefore, are expected to have 
little effect on the $W$ boson mass extracted. While initial 
state photon radiation increases the cross section by
$0.9\%$, weak one-loop corrections almost completely cancel the
initial state photonic corrections. The complete ${\cal O}(\alpha)$ 
initial state EW
corrections reduce the leading order (LO) cross section by about 
0.1\%. Initial and final state photon radiation interfere very little. 
The interference effects are uniform and have essentially no 
effect on the $M_T$ distribution. Final state photon radiation 
changes the shape of the transverse mass distribution
and reduces the LO cross section by up to $1.4\%$ in the
$W$ resonance region. Weak corrections again have no influence on the 
lineshape, but reduce the cross section by about 1\%. The $W$ mass
obtained from the $M_T$ distribution including the full EW one-loop
corrections is expected to be several MeV/c$^2$ smaller than that 
extracted employing the approximate calculation of Ref.~\cite{BK}.

Since final state photon radiation introduces a significant shift in
the $W$ mass, one also has to worry about multiple photon radiation. A
calculation of $p\bar p\to\mu\nu\gamma\gamma$~\cite{BHSZ} which includes
all initial and final state radiation and finite muon mass effects shows
that approximately 0.8\% of all $W\to\mu\nu$ events contain two photons
with $E_T(\gamma)>0.1$~GeV (the approximate tower threshold of the
electromagnetic calorimeters of CDF and D\O) and $\Delta R(\gamma,
\gamma)>0.14$. This suggests that the additional shift in $M_W$ from 
multiple photon radiation may not be negligible if one aims at a
measurement with a precision of ${\cal O}(10~{\rm MeV/c}^2)$. 

\subsubsection{LHC}

At the LHC, the cross section for $W$ production is about a factor~4
larger than at the Tevatron. During the first year of operation, it is
likely that the LHC will run at a reduced luminosity of approximately
${\cal L}=10^{33}~{\rm cm}^{-2}\,{\rm s}^{-1}$, resulting in roughly $0.
9\times 10^7$ $W\to e\nu$ events with a central electron ($|\eta(e)|<1.
2$) and a transverse mass in the range $65~{\rm GeV/c}^2<M_T<100~{\rm
GeV/c}^2$. A similar number of $W\to\mu\nu$ events is expected.
Both LHC detectors, ATLAS~\cite{atlas} and CMS~\cite{cms}, will be able
to trigger on electrons and muons with a transverse momentum of
$p_T(\ell)>15$~GeV/c ($\ell=e,\,\mu$), and should be fully efficient for
$p_T(\ell)>20$~GeV/c. They are well-optimized for electron, muon and 
$\etmis$ detection. 

At ${\cal L}=10^{33}~{\rm cm}^{-2}\,{\rm s}^{-1}$, the average number of
interactions per crossing at the LHC is approximately $I_C=2$, which is
significantly smaller than what one expects at the Tevatron for the same
luminosity. A precision measurement of the $W$ mass at the LHC running
at a reduced luminosity, using the traditional transverse mass analysis,
thus seems feasible~\cite{KW}. 

QCD corrections to the transverse mass distribution at the LHC enhance
the cross section by 10 -- 20\% in the $M_T$ range which is normally
used to determine $M_W$.
This is illustrated in Fig.~\ref{FIG:NINETEEN}, where the LO and NLO QCD
transverse mass distribution is shown, together with the NLO
to LO differential cross section ratio. Here, a $p_T(\ell)>20$~GeV/c and a
$p\llap/_T>20$~GeV/c cut have been imposed, and the pseudorapidity of the
lepton is required to be $|\eta(\ell)|<1.2$. The slight change in the
shape of the $M_T$ distribution induced by the NLO QCD corrections is
due to the cuts imposed. 
\begin{figure}[h]
\leavevmode
\begin{center}
\resizebox{8.9cm}{!}{%
\includegraphics{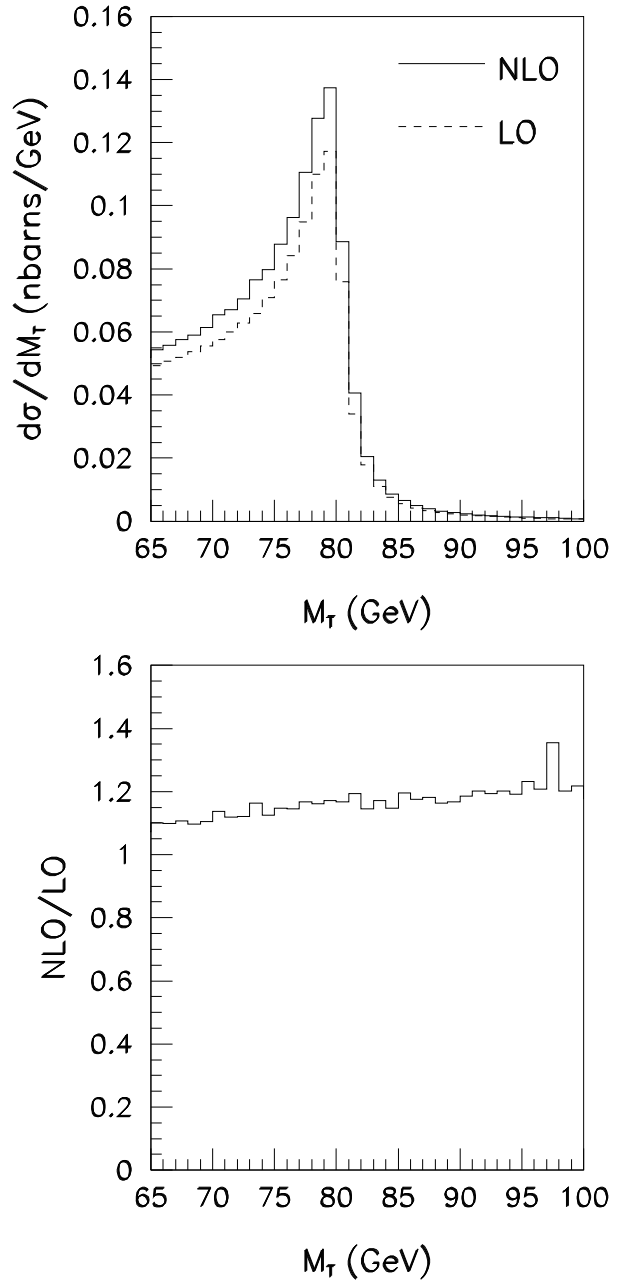}}
\end{center}
\caption{The LO and NLO QCD $W$ transverse mass distribution at the LHC.
Also shown is the NLO to LO differential cross section ratio as a
function of $M_T$.}
\label{FIG:NINETEEN}
\end{figure}

So far, no detailed study of the precision which one might
hope to achieve for $M_W$ at the LHC has been performed. For a
{\sl crude order of magnitude} estimate, one
can use the statistical and systematic errors of the current CDF and
D\O\ analyses~\cite{CDFWmass,D0Wmass}, and scale them by $\sqrt{I_C/N}$.
For an integrated luminosity of 10~fb$^{-1}$, one obtains~\cite{KW}:
\begin{equation}
\delta M_W \lappeq 15~{\rm MeV/c}^2.
\end{equation}
In order to see whether LHC experiments can perform a measurement of
$M_W$ which is significantly more precise than what one expects from
TeV33 or the NLC, a more detailed study which also considers other
quantities such as the transverse mass ratio of $W$ and $Z$
bosons~\cite{KG,srini} has to be carried out.

\subsubsection{Constraints on $M_H$ from $M_W$ and $M_t$}

The potential of extracting useful information on the Higgs boson mass
from a fit to the SM radiative corrections and a precise measurement of 
$M_W$ and $M_t$ is illustrated in 
Fig.~\ref{FIG:TWENTY}. Here we have assumed $M_t=176\pm 2$~GeV/c$^2$, 
$M_W=80.330\pm 0.010$~GeV/c$^2$, and $\alpha^{-1}(M_Z^2)=128.89\pm 0.05$.
\begin{figure}[t]
\leavevmode
\begin{center}
\resizebox{8.9cm}{!}{%
\includegraphics{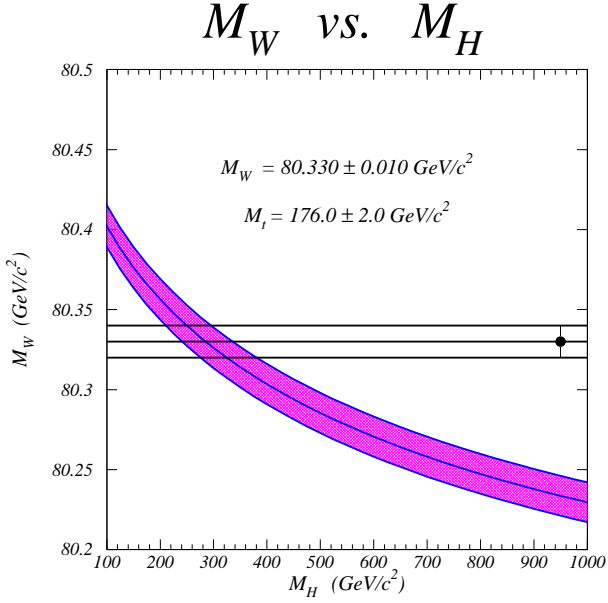}}
\end{center}
\caption{Predicted $W$ versus Higgs boson mass for $M_t=176\pm
2$~GeV/c$^2$. The theoretical predictions incorporate the effects of
higher order electroweak and QCD corrections.}
\label{FIG:TWENTY}
\end{figure}
Such a measurement would constrain the
Higgs boson mass to $M_H=285^{+65}_{-55}$~GeV/c$^2$. The corresponding
log-likelihood function is shown in Fig.~\ref{FIG:TWENTYONE}. 
\begin{figure}[t]
\leavevmode
\begin{center}
\resizebox{8.9cm}{!}{%
\includegraphics{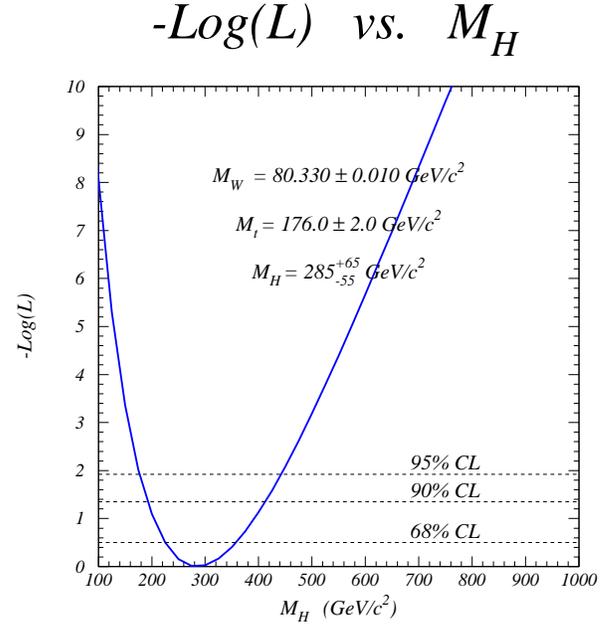}}
\end{center}
\caption{The negative log-likelihood function assuming $M_W=80.330\pm 0.
010$~GeV/c$^2$ and $M_t=176\pm 2$~GeV/c$^2$.}
\label{FIG:TWENTYONE}
\end{figure}
A measurement of the $W$ mass with a precision of $\delta
M_W=10$~MeV/c$^2$ and of the top mass with an accuracy of 2~GeV/c$^2$
thus translates into an indirect determination of the Higgs boson mass
with a relative error of about 
\begin{equation}
\delta M_H/M_H\approx 20\%.
\end{equation}
From a global analysis of all electroweak precision data one might then
expect $\delta M_H/M_H< 15\%$. 

For the precision of $M_t$ and $M_W$
assumed here, the theoretical error from higher orders and the 
uncertainty in the electromagnetic coupling constant
$\alpha(M_Z^2)$ become limiting factors for the accuracy which can be
achieved for $M_H$. Efforts to calculate higher order corrections and to
significantly improve the error on $\alpha(M_Z^2)$ beyond what one can 
expect from measurements at Novosibirsk, DAP$\Phi$NE, or BES, 
need increased emphasis from both experimentalists and
theorists in order to be able to achieve an ultimate relative precision
on $M_H$ better than about 15\%. 

\section{Summary and Conclusions}

In this report, we have highlighted some current high precision electroweak
measurements, and explored prospects for further improvements over
the next decade. The aim of precision electroweak measurements is to
test the SM at the quantum level, and to extract indirect
information on the mass of the Higgs boson. The confrontation of these
indirect predictions of $M_H$ with the results of direct searches for
the Higgs boson will be perhaps the most exciting development of the
next decade in the field of particle physics.

Although a global fit to all available precision electroweak data yields
$M_H=149_{-82}^{+148}$~GeV/c$^2$, the Higgs boson mass extracted
strongly depends on the input quantities used in the fit. Excluding a
particular observable which displays a statistically significant
deviation from the SM prediction, {\it e.g.} the SLD left-right
asymmetry, may easily increase the central value of $M_H$ by a
factor~4. One therefore has to conclude
that present data are not quite sufficient to obtain a stable estimate of
the Higgs boson mass. 

Results of future collider experiments are expected to drastically
change this situation. In these experiments one hopes to precisely 
determine three
observables which are key ingredients in obtaining reliable indirect
information on the Higgs boson mass:
\begin{itemize}

\item The uncertainty on the top quark mass is expected
to be reduced by at least a factor~3 in Tevatron and LHC experiments. At
the NLC or a $\mu^+\mu^-$ collider, a precision of a few hundred
MeV/c$^2$ may be possible.

\item It should be possible to reduce the error on
$\sin^2\theta_{eff}^{lept}$ by at least a factor two through
measurements of the left-right asymmetry at a luminosity upgraded SLC,
and the forward backward asymmetry in the $Z$ peak region at the
Tevatron and LHC.

\item The most profound improvement is likely to occur for the $W$ mass,
where a gain of a factor~5 seems to be within reach.
New strategies developed for extracting $M_W$ at hadron
colliders~\cite{KG,srini} will
make it possible to fully exploit the expected increase in integrated
luminosity at the Tevatron. 

\end{itemize}
From a measurement of $M_t$ with a precision of 2~GeV/c$^2$, and $M_W$
with an uncertainty of 10~MeV/c$^2$ alone it should be possible to
constrain $M_H$ within 20\%. 

As the electroweak measurements improve, the theoretical error from
higher orders and the uncertainty in
$\alpha(M_Z^2)$ will gradually become more and more important
limitations in the precision
which can be achieved. The determination of $\alpha(M_Z^2)$ is limited by
the knowledge of the photon hadron coupling at small momentum transfer.
An increased experimental and theoretical effort is needed to overcome
the present limitations in determining $\alpha(M_Z^2)$, and to calculate
higher order corrections to the electroweak observables.
\newpage
%
%%%%% References
%

\end{document}
%%% EOF